%% file: main.tex
\documentclass[sigconf,nonacm]{acmart}
\AtBeginDocument{%
  \providecommand\BibTeX{{%
    \normalfont B\kern-0.5em{\scshape i\kern-0.25em b}\kern-0.8em\TeX}}}
\usepackage{enumitem}
\usepackage{natbib}
\usepackage{tabularx}
\usepackage{subcaption} 
\usepackage{framed}

\begin{document}

\title{The Belief Update Gate: Separating Inertia from Learning in Human-AI Interaction}
\author{Shreyan Biswas}
\affiliation{%
  \institution{Delft University of Technology}
  \city{Delft}
  \country{The Netherlands}}
\email{s.biswas@tudelft.org}

\author{Alexander Erlei}
\affiliation{%
  \institution{University of Goettingen}
  \city{Goettingen}
  \country{Germany}}
\email{alexander.erlei@wiwi.uni-goettingen.de}

\author{Ujwal Gadiraju}
\affiliation{%
  \institution{Delft University of Technology}
  \city{Delft}
  \country{The Netherlands}}
\email{u.k.gadiraju@tudelft.nl}

\input{sections/00_abstract}



\maketitle

\input{sections/01_introduction}
\input{sections/02_background}
\input{sections/03_method}
\input{sections/04_results}
\input{sections/05_discussion}
\input{sections/06_conclusion}
\begin{acks}

We used generative AI tools, including ChatGPT, to assist with language polishing, grammar correction, and clarity-oriented revision of author-written text. We also used generative AI tools to assist with utility code for analysis scripts and plotting. All text, code, analyses, figures, interpretations, and citations were reviewed, edited, and validated by the authors, who take full responsibility for the content of this work.
\end{acks}
\bibliographystyle{ACM-Reference-Format}
\bibliography{references}
\input{sections/07_appendix}
\end{document}

%% file: sections/00_abstract.tex
\begin{abstract}
Repeated human--AI interaction is often analyzed through pooled belief-updating slopes: users observe AI successes and failures, revise reported beliefs in the feedback-consistent direction, but appear conservative on average. We show that such averages can obscure an important distinction between whether an elicited belief report changes at all and how it changes conditional on movement. We refer to this measurement-aware decomposition as the \textit{belief update gate}. Reanalyzing a multi-task human--AI decision-making dataset with 240 participants, 7{,}200 trials, and three task domains, we find substantial non-movement in reported beliefs: 67.3\% of trial-level belief changes are exactly zero, and 76.4\% are smaller than five percentage points. Separating non-moving from moving reports changes the descriptive interpretation of pooled conservatism: the within-trajectory slope rises from 0.494 overall to 0.949 among rows with nonzero movement. Since this latter estimate conditions on observed movement, we interpret it as a descriptive decomposition rather than as evidence of a near-Bayesian latent learning process. Complementary hurdle style analyses (i.e., modeling zero vs. non-zero changes before predicting update magnitude) show that the absolute discrepancy between feedback and entering belief predicts whether a report changes, while the signed feedback discrepancy predicts the direction and magnitude of change among reports that move. Importantly, observed non-movement does not distinguish genuine latent belief inertia from small unexpressed updates, rounding, or other reporting processes. These findings show that calibration analyses of repeated human--AI interaction should distinguish visible non-movement in elicited belief reports from updating conditional on movement rather than treating reported beliefs as a single continuous updating process.
\end{abstract}

%% file: sections/01_introduction.tex
\section{Introduction}

AI systems are increasingly embedded in everyday workflows. Rather than encountering AI in isolated decisions, people now interact with the same or similar systems across heterogeneous tasks, including information seeking, writing, programming, planning, and decision support \cite{handa2025economictasksperformedai,chang2024survey,lawless2024want,bianchini2025exploring}. In this setting, the key human factor is no longer just whether users accept or reject one AI recommendation. It is how experience with the system accumulates across tasks and shapes future reliance.

This shift makes repeated experience consequential. Each observed success or failure can change what users expect from the system: when it is useful, when it is reliable, and when it is worth deferring to. Prior work shows that effective human--AI teaming depends on users' mental models of AI capabilities and error boundaries, and that trust and reliance develop over time through prior interaction experience \cite{tolmeijer2021second,bansal2019beyond,kahr2024understanding,lu2024mix}. A user who sees an AI system fail in one task may become more skeptical in later tasks; a user who sees it perform well may carry forward higher expectations, even when the next task differs substantially. Understanding human--AI interaction, therefore, requires studying not only immediate reliance decisions, but also how reported expectations about AI performance evolve through repeated experience.

One way to study belief revision in behavioral decision research is to estimate average responses to new information, such as feedback, signals, or observed outcomes. In this tradition, researchers ask whether people move their beliefs in the direction implied by the evidence, and how large that movement is relative to a normative benchmark such as Bayesian updating \cite{edwards1982conservatism,hogarth1992order}. Recent human--AI work has begun to adapt this logic to repeated AI-assisted decision making. In the original analysis of the dataset reexamined here, \citet{biswas2026belief} modeled belief updating by comparing users' reported belief changes after AI successes and failures to a Bayesian benchmark.

This population-level approach is useful because it gives a compact answer to an important calibration question: on average, do users adjust their reported beliefs about AI accuracy in the direction suggested by feedback, and by how much? It also connects naturally to human--AI reliance, since appropriate reliance depends partly on whether users adjust expectations as evidence about system performance accumulates \cite{lu2021human,zhang2020effect,kahr2024understanding,ma2023who}. However, the same average can conceal qualitatively different patterns of observed behavior. An average updating slope can obscure a more basic question: \emph{does feedback produce a measurable change in the reported belief at all?} If many reports remain unchanged after feedback, then a pooled slope combines two components of observed updating behavior: the probability that a report moves and the size of the movement conditional on movement. In such cases, apparent conservatism may partly reflect the composition of moving and non-moving reports rather than uniformly weak adjustment across observations.

This distinction is especially important because reported beliefs are elicited measurements, not direct observations of latent belief states. A flat report may reflect genuine latent non-updating, but it may also arise from a small latent change that does not cross a reporting threshold, rounding or focal use of the response scale, satisficing, fatigue, or other features of the elicitation process. Consequently, observed non-movement does not by itself identify latent belief inertia.

Two-part, hurdle, and double hurdle approaches provide a natural framework for separating the occurrence of adjustment from its magnitude, conditional on adjustment \cite{henckel2022belief}. Building on this general perspective, we examine whether repeated reported belief updating in human--AI interaction is more informatively characterized by distinguishing two observational components: first, whether feedback produces a measurable change in the elicited belief report; and second, how the report changes conditional on movement. We refer to this measurement-aware decomposition as the ``\textit{belief update gate}.''
We study this question by reanalyzing the experimental data from \citet{biswas2026belief}. The dataset contains 240 participants, 720 participant-task blocks, and 7{,}200 trial level observations from a repeated human--AI decision making experiment across three task domains: grammar, travel planning, and visual question answering. On each trial, participants observed an AI recommendation, decided whether to rely on it, received feedback about whether the AI was correct, and then reported their belief about the AI's future correctness. This design allows us to characterize visible movement and non-movement in repeated belief reports, examine updating conditional on movement, and explore how these reported-belief dynamics relate to downstream delegation.

This paper makes two main contributions and provides one supporting robustness analysis:

\begin{enumerate}[leftmargin=*,itemsep=2pt]

\item \textbf{A measurement-aware reanalysis of repeated human--AI belief updating.}
We show that separating visible non-movement from movement changes the descriptive interpretation of average conservative updating. Rather than implying uniformly weak adjustment across trials, the pooled estimate combines a large mass of non-moving belief reports with a smaller set of reports that move more strongly in the feedback-consistent direction, revealing how movement frequency and conditional adjustment jointly shape the aggregate estimate.

\item \textbf{An application of a two-part perspective to reported belief dynamics in human--AI interaction.}
Building on established hurdle style approaches, we use the term ``belief update gate'' to distinguish whether feedback produces measurable movement in an elicited belief report from how the report moves conditional on movement. This framing makes explicit that pooled calibration patterns can arise from different combinations of visible non-movement and conditional updating.

\item \textbf{Supporting robustness analyses for heterogeneous reported-belief trajectories.}
We assess whether the main measurement pattern is stable across alternative movement thresholds, predictor audits, and trajectory level model comparisons. These analyses characterize recurring patterns of visible non-movement, conditional updating, and heterogeneity across reported belief trajectories.
\end{enumerate}

The intended contribution is empirical and conceptual. Our results show that average updating estimates can conceal consequential heterogeneity in reported belief dynamics. Rather than asking only whether users update their beliefs ``too little'' on average after repeated AI exposure, we ask which reports move, what kinds of evidence are associated with movement, and how reports change conditional on movement. This perspective provides a more granular account of reported belief updating in repeated human--AI interaction and a basis for examining how these dynamics relate to subsequent reliance.

%% file: sections/02_background.tex
\section{Background and Literature Review}
\label{sec:background}

\subsection{Human Computation, Crowdsourcing, and AI-assisted Workflows}

Human computation research has increasingly moved from treating people as interchangeable labelers \cite{dawid1979maximum,diaz2022crowdworksheets} toward studying systems in which humans and AI collaborate as adaptive partners \cite{bansal2019beyond,mcgrath2025collaborative,zhao2025role,oppenlaender2025quo}. This shift matters for the present paper because repeated belief formation is 
a complex problem: task instructions, feedback policies, incentives~\cite{kaur2026incentive,holstein2026thinking}, and workflow structure shape whether users notice evidence, interpret it as diagnostic~\cite{salimzadeh2024dealing}, and express belief change in AI-assisted decision-making~\cite{mason2009financial,shaw2011designing,salimzadeh2024doubt,nouri2023supporting,dutta2024unveiling,he2023knowing,banuqitah2024exploratory,erlei2022s,erlei2024understanding,erlei2026life}.

This holistic perspective motivates a cautious interpretation of belief reports in repeated AI use. Apparent non-movement should not be interpreted immediately as a pure cognitive failure to learn. It may also reflect the elicitation setting~\cite{ahn2024impact,salimzadeh2024dealing}: whether feedback is diagnostic, whether small belief changes are meaningful for the task, and whether the response scale makes such changes visible. The current paper, therefore, treats reported belief updating as part of the human--AI workflow rather than as a direct readout of a latent mental state.

\subsection{Trust, Reliance, and Repeated Interaction in Human--AI Decision Making}

Trust and reliance are central constructs in human--AI decision making, but they are not static properties of a user or a system~\cite{gadiraju2025enterprising,mehrotra2024systematic}. In repeated interaction, users encounter a sequence of recommendations, observe successes and failures, and gradually form expectations about when the AI is likely to help or mislead them. This makes reliance a dynamic process: people not only decide whether to accept a particular recommendation, but also adapt future-oriented beliefs about the system, which can guide later reliance~\cite{tolmeijer2021second,kahr2024understanding}.

\citet{kocielnik2019will} show that expectation management matters because users' willingness to accept imperfect AI depends on how anticipated errors are framed. \citet{zhang2020effect} demonstrate that confidence displays can calibrate trust, but that calibration alone does not guarantee superior team performance. \citet{lu2021human} show that when performance feedback is limited, people adopt heuristics that can produce both over- and under-reliance. \citet{he2025conversational} showed that seamless conversational interactions geared to understand AI advice can amplify over-reliance. \citet{cao2022understanding} pair outcome measures with process traces such as gaze. Related work argues that appropriate trust depends not only on AI correctness likelihood but also on estimated human correctness likelihood \cite{ma2023who}, while dynamic models of trust and reliance show that observed reliance behavior can arise from latent affective and behavioral states \cite{li2023modeling}.

More recent work emphasizes both temporal development and heterogeneity. \citet{kahr2024understanding} find that model accuracy and prior interaction history matter strongly for reliance. \citet{lu2024mix} show that population-level behavior is better described as a combination of distinct decision strategies than by a single average process, and \citet{cooper2025trust} argue that trust updates track users' expectations rather than raw outcomes alone. Recent reviews further argue that the field still lacks consensus on how to measure appropriate reliance dynamically and objectively, especially over repeated interactions rather than isolated advisory decisions \cite{raees2026people,raees2026trust}. Taken together, this literature supports the premise that repeated human--AI interaction should be modeled dynamically and heterogeneously, but it leaves open where heterogeneity enters the belief--reliance process. The present paper focuses on one such layer: reported belief updating. We ask whether repeated evidence becomes measurable movement in elicited belief reports, how reports move conditional on movement, and how these reported belief dynamics relate to delegation.

\subsection{Behavioral Models of Belief Updating}

Average belief updating can conceal substantial heterogeneity in how people respond to evidence. Most directly, \citet{henckel2022belief} develop a double hurdle model of sequential belief adjustment in which the first hurdle captures whether a stated belief changes and the second captures the extent of adjustment conditional on change. Their controlled experiment further relates these components to accumulated evidence, inattention, complexity, and deviations from Bayesian updating.

We build on this perspective in repeated human--AI interaction, where users report expectations about AI correctness across successive encounters. We examine how prevalent non movement in these reports contributes to pooled estimates of conservative updating, how movement varies with trial-level feedback, and how reported-belief dynamics relate to downstream delegation.

More broadly, behavioral research has documented several departures from homogeneous Bayesian updating. These include conservative adjustment \cite{edwards1982conservatism}, order and processing effects \cite{hogarth1992order}, confirmatory bias \cite{rabin1999first}, asymmetric responses to good and bad news \cite{eil2011good}, and misperceptions of sample informativeness \cite{benjamin2016model}. Together, these accounts motivate examining repeated belief reports as potentially heterogeneous in both whether they move and how they move conditional on movement.

%% file: sections/03_method.tex
\section{Method}

\subsection{Original Experiment and Reanalysis Overview}
\label{method:original-exp}
We reanalyze data\footnote{https://osf.io/65s2u/files/osfstorage/6a26e999df27f810336d825c} from a repeated human--AI decision-making experiment in which participants interacted with AI assistance across three task domains: grammar, travel, and visual question answering (VQA) (Appendix~\ref{app:interface}). Each participant completed 10 trials per task. On each trial, participants saw an AI answer, chose whether to use the AI, observed whether the AI was correct, and then reported their belief about the AI's future correctness on a 0--100 scale. Participants also reported self-confidence. Before each task, they provided counterfactual prior judgments about what they would believe after a single correct or incorrect AI outcome.

The raw valid-session file contains 240 sessions passing the study-level validity checks; details of these checks are reported in the original paper~\cite{biswas2026belief}. The resulting balanced dataset contains 240 participants, 720 participant-task blocks, and 7{,}200 trial observations. Unless otherwise noted, analyses use this full trial panel.

\subsection{Derived Variables}
For each participant $i$, task $j$, and trial $t$, we construct:
\begin{itemize}[leftmargin=*,itemsep=2pt]
    \item $y_{ijt}$, an indicator for whether the AI was correct on trial $t$;
    \item $b_{ijt}^{lag}$, the belief entering trial $t$, scaled to $[0,1]$;
    \item $b_{ijt}^{post}$, the belief reported after feedback on trial $t$, scaled to $[0,1]$;
    \item $\Delta b_{ijt}=b_{ijt}^{post}-b_{ijt}^{lag}$, the observed belief update;
    \item $d_{ijt}$, an indicator for delegation to the AI;
    \item $c_{ijt}^{lag}$, the participant's lagged self-confidence.
\end{itemize}
For the first trial of a task, $b_{ijt}^{lag}$ is the task-opening AI belief. For later trials, it is the previous trial's post-feedback belief. 

\subsection{Bayesian Benchmark and Hybrid Prior Rule}

Several candidate models compare observed belief updates to the Bayesian benchmark used in the original study. Following \citet{biswas2026belief}, we use a Beta--Bernoulli updating process for binary AI correctness feedback. If a participant enters task $j$ with prior mean $p_{ij0}$ and effective prior sample size $S_{ij}$, then after $c$ AI successes and $f$ AI failures the benchmark posterior mean is
\begin{equation}
p_{ij}(c,f)=\frac{S_{ij}p_{ij0}+c}{S_{ij}+c+f}.
\end{equation}

We use the original paper's hybrid $S=10$ specification as the canonical source panel for benchmark-dependent model comparison. In this specification, participant-task blocks with coherent counterfactual-prior responses use the estimated $S_{ij}$ from the task-level counterfactual judgments described above in Section~\ref{method:original-exp}. Otherwise valid blocks use a fixed fallback value of $S=10$, up to the fallback cap used in the original benchmark construction.

Of the 720 participant-task blocks, 440 use estimated counterfactual-prior values and 216 use the $S=10$ fallback. The remaining 64 blocks are not assigned trajectory-level model labels because their counterfactual-prior estimates did not pass the benchmark-construction checks and the fallback allocation was already capped. These blocks are retained in analyses that use only observed beliefs, feedback, confidence, and delegation.

We also repeat the benchmark-dependent analyses under the strict, lenient, hybrid $S=5$, and hybrid $S=20$ specifications to check whether the headline pattern depends on the source-panel choice (Appendix~\ref{app:source-panel-robustness}).

\subsection{Modeling Framework}

The analysis is organized around a distinction between latent belief formation and the measurement process through which beliefs are reported. Rather than treating each 0--100 report as a transparent readout of an underlying belief, we separate two questions: whether feedback produces measurable movement in the elicited belief report, and how the report changes conditional on movement. We also summarize how belief-report movement relates to delegation, treating delegation as a downstream reliance behavior rather than as the main explanation for the belief update gate.

This framework is measurement-aware. An observed flat belief report sequence is not point-identifying: it could reflect genuine latent inertia, a latent update too small to cross a reporting threshold, or coarse/focal use of the 0--100 scale. The analyses, therefore, identify visible non-movement in reported beliefs, not latent non-learning with certainty.

\subsection{Slope-Decomposition Analysis}

To examine whether the original pooled conservative-updating estimate is driven by visible non-movement, we recompute the within-trajectory slope between observed belief updates and the Bayesian benchmark across nested analysis subsets. We estimate the slope for all classified rows, after excluding \texttt{no\_update} winner trajectories, within trajectories containing at least two nonzero belief movements, and among rows with nonzero belief movement only. We compute confidence intervals using participant-level bootstrap resampling. It is worth noting that this analysis is descriptive: the moved-row subset conditions on observed movement and is used to diagnose composition in the pooled estimate, and not to estimate a causal learning rate.

\subsection{Update-Gate Regression Analysis}

We estimate a two-part, hurdle-style regression analysis on the full 7{,}200-trial panel. We use the term ``hurdle-style'' because the analysis separates whether a belief report crosses a movement threshold from the size and direction of movement conditional on crossing that threshold. 

For the first part, we define a movement indicator
\[
M_{ijt}^{(0)}=\mathbb{1}\left(|\Delta b_{ijt}|>10^{-12}\right).
\]
For robustness thresholds, we define
\[
M_{ijt}^{(\tau)}=\mathbb{1}\left(|\Delta b_{ijt}|\ge \tau\right),
\quad \tau\in\{.01,.02,.05\}.
\]

We estimate participant-clustered logistic regressions predicting $M_{ijt}^{(\tau)}$. The key predictor is the \texttt{absolute feedback gap}, $|y_{ijt}-b^{lag}_{ijt}|$, which captures how discrepant the observed AI feedback is from the participant's lagged belief entering the trial. The model also includes lagged belief, lagged self-confidence, task domain, and trial ordinal. Robustness specifications additionally control for trajectory-level updater family, with unclassified trajectory blocks retained as a separate category. We repeat this analysis for exact movement and for movements of at least 1, 2, and 5 percentage points (Appendix~\ref{app:update-gate-thresholds}).

For the second part, we restrict the data to rows where $M_{ijt}^{(\tau)}=1$ and estimate participant-clustered linear regressions predicting signed belief movement, $\Delta b_{ijt}$. The key predictor is the signed feedback gap, $y_{ijt}-b^{lag}_{ijt}$, with the same controls. This conditional model asks whether moved reports change in the feedback-consistent direction and how strongly they move once movement occurs.

\subsection{Predictors of Belief Report Movement}

Because the update-gate account raises the question of why reports do not move, we ask whether report movement is associated with participant characteristics, task context, performance, or local feedback structure.

We estimate two sets of regressions. First, at the participant-task level, we define an indicator for whether \texttt{no\_update} is the BIC winning trajectory model, and separately analyze all-zero movement trajectories. We estimate task only, objective accuracy, behavioral, and full pre-survey logistic specifications. The full pre-survey specification includes opening belief, mean self-confidence, final user accuracy, first AI feedback, task domain, task position, and pre-survey covariates, including AI literacy (MAILS) \cite{carolus2023mails}, need for cognition (NCS-6) \cite{lins2020very}, and trust in automation propensity (TiA) \cite{korber2018theoretical}. A separate objective-accuracy specification tests whether realized AI accuracy predicts inertial classification. Together, these models ask whether non-moving belief-report trajectories are concentrated among particular users, tasks, or starting conditions.

Second, at the trial level, we model whether the belief report moves on a given trial. We define movement as
\[
M_{ijt}=\mathbb{1}\left(|\Delta b_{ijt}|>10^{-12}\right),
\]
and estimate participant-clustered logistic regressions predicting $M_{ijt}$. Predictors include feedback discrepancy, AI feedback valence, lagged belief, lagged self-confidence, observed user correctness, task domain, trial ordinal, and pre-survey covariates. This model asks whether local evidence and trial context predict whether the report moves.

We also summarize report movement by delegation choice. We treat this comparison descriptively because delegation is a downstream reliance behavior rather than a causal explanation for the update gate. All regressions report participant-clustered standard errors and two-sided $p$-values for planned regression coefficients and group contrasts (Appendix~\ref{app:inertia-predictor-audit}).

As an additional participant-level diagnostic, we compare participants who never show visible report movement across all 30 trials with participants who move at least once. This comparison summarizes mean absolute feedback discrepancy, opening and terminal calibration error relative to realized task accuracy, and inferred or fallback effective prior strength.

\subsection{Trajectory-Level Model Comparison}

We compare 18 interpretable candidate models to characterize short participant-task belief-report trajectories. The candidate set covers theoretically meaningful alternatives: visible inertia; Bayesian and weighted-Bayesian benchmarks; recency and memory; thresholding, stickiness, and coarse expression; asymmetric responses to AI successes and failures; and temporal-regime models such as first impressions or changepoints. Appendix~\ref{app:model-equations} gives the model equations, and Appendix~\ref{app:model-slate} lists the implemented model slate.

For interpretation, we distinguish between full-path trajectory models and conditional update-rule models. Full-path models start from the task-opening belief and generate a predicted 10-trial belief path from the observed AI correctness sequence. Conditional update-rule models instead condition on the participant's observed lagged belief report on each trial and predict the next belief change. Because they condition on intermediate observed reports rather than generating them, we treat conditional update-rule models as robustness checks rather than as the main trajectory-level comparison. Further fitting details and model-selection criteria are reported in Appendix~\ref{app:model-fit}; conditional update-rule winners are reported in Appendix~\ref{app:all-candidate-winners}.

\subsection{Recovery and Identifiability Checks}

Each participant-task trajectory contains only 10 trials, so different candidate models can generate similar reported-belief paths. We therefore use design-matched recovery simulations to assess which model distinctions are recoverable under the same short-panel design~\cite{palminteri2017importance,wilson2019ten}. For each generating model, we simulate 25 trajectories under the observed 10-trial design and refit the candidate model set using both BIC and AICc. Self-recovery is defined as the proportion of simulated trajectories for which the generating model is selected as the best-fitting model. These checks indicate which trajectory-level distinctions can be interpreted reliably under the available design.

%% file: sections/04_results.tex
\section{Results}

The results decompose pooled conservative updating into visible non-movement and conditional movement. We first document the large mass of zero and near-zero belief-report changes, then show how the pooled slope changes when non-moving reports are separated from active-movement rows. We then examine the two-part pattern directly, audit predictors of belief-report movement, and compare short belief-report trajectories using interpretable model families.

\subsection{Descriptive Evidence for Visible Non-movement}

\begin{figure}[t]
\centering
\includegraphics[width=\linewidth]{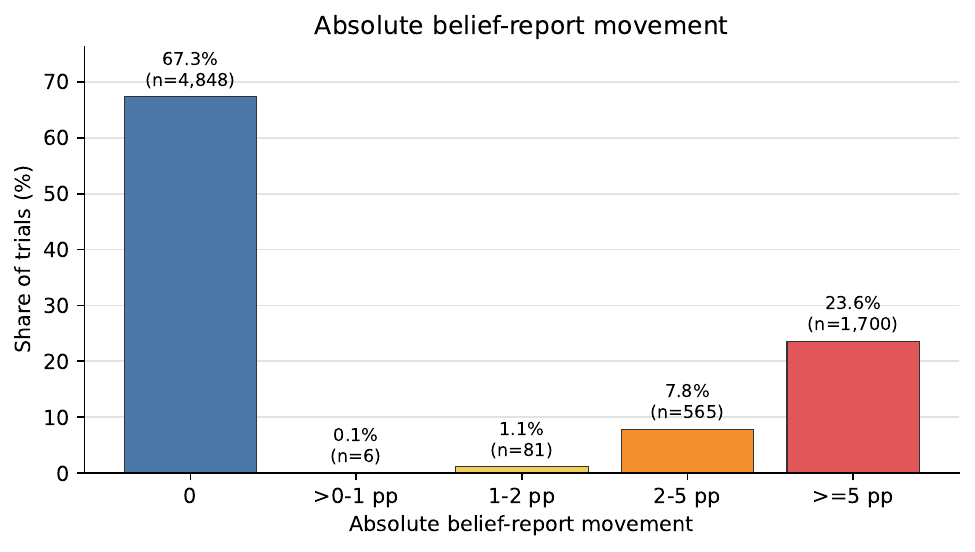}
\caption{Distribution of absolute trial-level belief-report movement. Two-thirds of trial-level reports show exactly zero movement, while only 23.6\% move by at least five percentage points.}
\Description{Bar chart showing that 67.3 percent of trials have zero belief movement, 0.1 percent have greater than zero but less than one percentage point, 1.1 percent have one to two percentage points, 7.8 percent have two to five percentage points, and 23.6 percent have at least five percentage points.}
\label{fig:belief-movement-bins}
\end{figure}

The strongest descriptive fact is the prevalence of non-movement. Across the full 7{,}200-trial behavioral panel, 67.3\% of belief changes are exactly zero, and 76.4\% are smaller than five percentage points (Figure~\ref{fig:belief-movement-bins}). With such a large mass of non-movement, a moderate average slope may partly reflect the composition of non-moving and active-movement rows.

The trajectory-level evidence is consistent with this descriptive pattern. In the canonical primary-generative comparison, \texttt{no\_update} is the modal winner, accounting for 233 of 656 classified participant-task trajectories (35.5\%; Table~\ref{tab:winners}). Together, the trial- and trajectory-level results show that non-movement constitutes a substantial component of the observed belief-report data.

\subsection{Why Does the Pooled Slope Still Look Conservative?}

\begin{figure}[t]
\centering
\includegraphics[width=\linewidth]{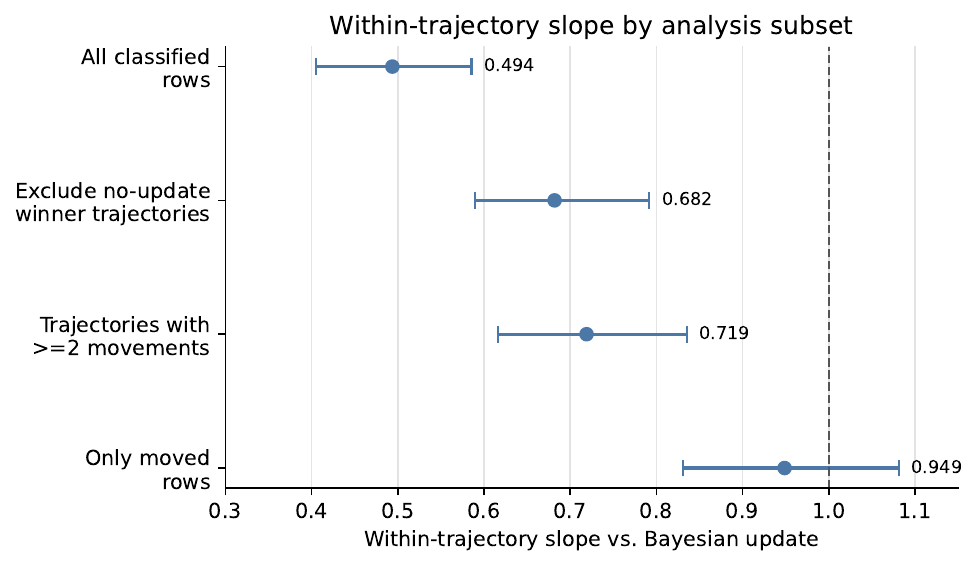}
\caption{Participant-bootstrap slope decomposition. The estimated within-trajectory slope rises as zero-movement trajectories and rows are separated from active movement. Intervals are 95\% participant-bootstrap confidence intervals.}
\Description{Point-and-interval plot showing the slope estimate increasing from 0.494 for all classified rows to 0.682 after excluding no-update winner trajectories, 0.719 among trajectories with at least two movements, and 0.949 among moved rows only.}
\label{fig:slope-decomposition}
\end{figure}

The subset diagnostics explain (Figure~\ref{fig:slope-decomposition}) how these non-moving reports coexist with the original pooled within-trajectory slope of 0.4937. When \texttt{no\_update} winner trajectories are removed, the slope rises from 0.494 (participant-bootstrap 95\% CI [.405,.586]) to 0.682 [.590,.792]. Restricting the sample to trajectories with at least two nonzero updates raises the slope to 0.719 [.617,.835]. Looking only at rows with nonzero belief changes yields a slope of 0.949 [.831,1.081]. Because this last subset conditions on observed movement, it is diagnostic rather than causal: the pooled conservative slope is real, but it partly reflects a mixture of zero-movement and active-movement observations.

This pattern illustrates why separating visible movement from conditional movement is informative: the pooled conservative-Bayesian estimate combines the frequency of report movement with the magnitude of movement when it occurs.

\subsection{A Direct Update-Gate Diagnostic}

In a trial-level logistic model of visible report movement, the absolute gap between current belief and feedback strongly predicts whether a report changes ($\hat{\beta}=1.332$, participant-clustered $z=10.09$, $p=5.88\times10^{-24}$, odds ratio $=3.79$, $n=7{,}200$). Among the 2{,}352 rows with nonzero movement, the signed feedback gap predicts signed update magnitude ($\hat{\beta}=.192$, participant-clustered $t=19.80$, $p=2.80\times10^{-87}$, $R^2=.430$). The pattern is stable to stricter movement thresholds of 1, 2, and 5 percentage points (Appendix~\ref{app:update-gate-thresholds}). This supports the narrower claim that repeated reported-belief updating is better characterized by separating visible movement from movement conditional on change than by applying a single slope uniformly across observations.

\subsection{Predictors of Belief-Report Movement and Non-Movement}

The predictor audit asks whether \texttt{no\_update} trajectory classification is associated with task, confidence, performance, or pre-survey covariates. At the participant-task level, the answer is mostly no. In the full task-level model, opening belief, mean self-confidence, final user accuracy, first AI feedback, AI literacy, need for cognition, and trust in automation propensity do not separately predict \texttt{no\_update} classification; all have cluster-robust $p>.20$. Task domain is also not significant at the participant-task level (joint Wald $p=.812$). Task position is significant (joint Wald $p=.029$), but we interpret this as an order-associated pattern rather than as a causal estimate.

A separate cross-task consistency check shows that \texttt{no\_update} classification is partly participant-stable but not universal: participants with more \texttt{no\_update} winners in their other task blocks are more likely to receive the same classification in the current block, but many receive this classification in only one task and not in others (Appendix~\ref{app:inertia-consistency}).

The participant-level never-mover diagnostic does not indicate that never-movers simply encountered less discrepant feedback. Never-movers and ever-movers have nearly identical mean absolute feedback gaps (.443 versus .446). Never-movers start only slightly closer to objective AI accuracy (.295 versus .305 mean absolute calibration error), but by task end they are less calibrated than ever-movers (.295 versus .211). Never-movers also have somewhat higher inferred or fallback prior strength, partly driven by greater use of the fixed $S=10$ fallback, so this difference should be interpreted cautiously.

At the trial level, the clearest predictors are local. Larger feedback discrepancy increases the odds of visible report movement (odds ratio $=1.32$, $p=1.81\times10^{-12}$), while higher lagged belief lowers the odds of movement (odds ratio $=.72$, $p=7.29\times10^{-6}$). Observed user correctness is positively associated with visible movement (odds ratio $=1.28$, $p=6.26\times10^{-10}$), and AI-correct feedback is associated with lower odds of movement after controlling for absolute feedback gap (odds ratio $=.88$, $p=.00077$). This residual pattern is consistent with asymmetric sensitivity to AI failures, ceiling effects among high-belief reports, or coarse reporting; it should not be read as uniquely identifying one psychological mechanism.

\begin{figure}[t]
    \centering
    \includegraphics[width=\linewidth]{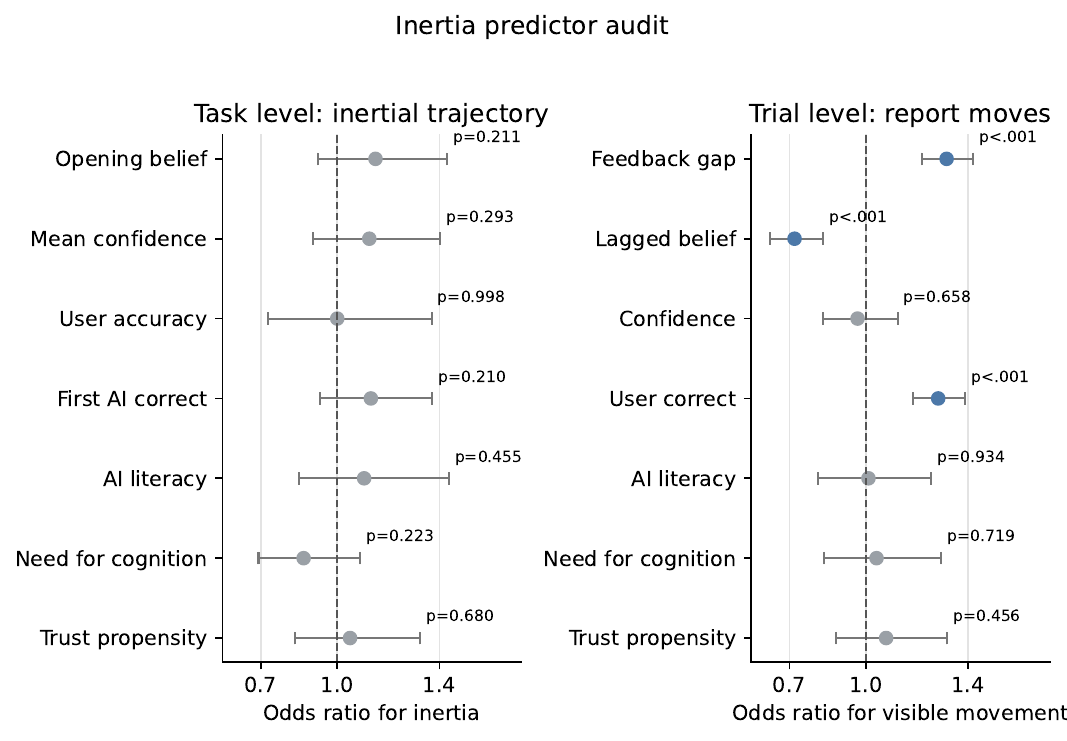}
    \caption{Predictor audit for visible inertia and visible movement. The left panel models participant-task \texttt{no\_update} trajectory classification, so odds ratios above 1 indicate greater odds of \texttt{no\_update} being selected. The right panel models trial-level visible report movement, so odds ratios above 1 indicate greater odds that a report moves. Points are odds ratios for standardized continuous predictors, intervals are participant-clustered 95\% confidence intervals, and labels show cluster-robust \(p\)-values.}
    \Description{Two-panel odds-ratio plot. The task-level panel shows opening belief, mean confidence, user accuracy, first AI correctness, AI literacy, need for cognition, and trust propensity all with confidence intervals crossing one and non-significant p-values. The trial-level panel shows feedback gap and user correctness above one with p less than .001, lagged belief below one with p less than .001, and confidence plus pre-survey traits with non-significant p-values.}
    \label{fig:inertia-predictor-audit}
\end{figure}

Figure~\ref{fig:inertia-predictor-audit} summarizes the selected standardized predictors from these two audit models. The categorical task and trial-ordinal tests are reported in text because their effects are joint contrasts rather than single odds ratios.

Trial ordinal is significant in the categorical trial-level model (joint Wald $p=1.33\times10^{-5}$), with visible update shares front-loaded: .37, .36, and .37 in trials 1--3, then about .29--.33 thereafter. Because trial ordinal is entangled with the realized feedback sequence and evolving belief levels, we treat this as evidence of temporal structure rather than proof that time itself causes inertia.

Visible update shares are similar on non-delegated and delegated trials (.33 versus .30). Thus, belief-report movement differs little by delegation choice in these descriptive comparisons.

\subsection{Trajectory-level Model Comparison}
Table~\ref{tab:winners} summarizes the canonical BIC winner counts for the hybrid $S=10$ source panel over 656 participant-task trajectories, restricting the main table to primary trajectory-generating models. Conditional update-rule models are reported separately in Appendix~\ref{app:all-candidate-winners} because they condition on observed lagged reports that the model itself did not generate.

The two most frequent winners summarize different forms of observed heterogeneity. The \texttt{no\_update} model summarizes visibly flat trajectories, whereas \texttt{good\_bad\_news} captures asymmetric responses to AI successes and failures. The remaining threshold, sticky, and coarse-reporting models capture cases where reports change only after sufficiently large evidence, remain anchored to prior reports, or move on a coarser response grid.

Three descriptive patterns stand out. First, visible non-movement is the modal trajectory motif when only primary generative models compete. Second, \texttt{good\_bad\_news} is the most frequent active-updating competitor. Third, threshold, sticky, and coarse-reporting models also win for subsets of trajectories, indicating that several qualitatively different model families can account for portions of the observed heterogeneity. These distinctions are interpreted in light of the recovery results below.

Design-matched recovery simulations temper these labels (Table~\ref{tab:recovery}). BIC self-recovery is 60\% for \texttt{no\_update}, 48\% for \texttt{threshold\_bayes}, and 40\% for \texttt{good\_bad\_news}, but substantially lower for the remaining families (4--16\%). AICc produces a similar pattern. We therefore interpret winner labels as descriptive motifs rather than definitive participant-level diagnoses.

\begin{table}[ht]
\centering
\caption{Self-recovery shares under BIC and AICc in the design-matched recovery
simulation.}
\label{tab:recovery}
\scalebox{0.8}{
\begin{tabular}{lrr}
\toprule
Generating model & BIC self-recovery & AICc self-recovery \\
\midrule
\texttt{no\_update} & 60\% & 64\% \\
\texttt{threshold\_bayes} & 48\% & 52\% \\
\texttt{good\_bad\_news} & 40\% & 44\% \\
\texttt{sticky\_weighted\_bayes} & 16\% & 12\% \\
\texttt{divisible\_weighted\_bayes} & 8\% & 8\% \\
\texttt{discounted\_weighted\_bayes} & 4\% & 4\% \\
\texttt{coarse\_weighted\_bayes} & 4\% & 4\% \\
\bottomrule
\end{tabular}}
\end{table}

%% file: sections/05_discussion.tex
\section{Discussion}

\subsection{From Average Conservatism to a \textit{Belief Update Gate} Account}

This reanalysis adds nuance to how the original conservative-updating result should be interpreted~\cite{biswas2026belief}. The original study showed that, on average, users revised their reported beliefs in the feedback-consistent direction but by less than a Bayesian benchmark would predict. That aggregate result remains important: it shows that users are not ignoring evidence entirely and that reported beliefs are behaviorally meaningful for reliance. This connects to classic work on belief revision, where people often update conservatively relative to Bayes~\cite{edwards1982conservatism}, and where updating can depend on order, processing strategy, and task structure~\cite{hogarth1992order}. In human--AI decision making, the same issue matters because appropriate reliance depends on whether users adjust expectations as evidence about system performance accumulates~\cite{zhang2020effect,lu2021human,kahr2024understanding,ma2023who,tolmeijer2021second}.

However, the present analysis shows that a pooled updating slope is not a sufficient description of the process. Pooled conservatism combines two empirically distinct components: many elicited belief reports do not measurably move after feedback, while reports that do move are substantially more aligned with feedback than the pooled estimate suggests. A low average updating slope should therefore not automatically be interpreted as evidence that every user updates weakly on every trial. The same average can arise from non-movement, asymmetric responses to successes and failures~\cite{eil2011good}, and rounded, focal, or coarse probability reports~\cite{manski2010rounding,kleinjans2014rounding,jakobsen2026coarse}.

We describe this distinction as a belief update gate. The term does not imply that we have identified a discrete cognitive switch inside the user. Rather, it names a measurement-level problem in repeated belief data: before asking how strongly a belief report moves, we must first ask whether it moves at all. The results support this account in three ways. First, exact-zero and near-zero changes are common. Second, the slope decomposition shows that apparent conservatism is partly compositional: when non-moving reports are separated from moving reports, the estimated response to feedback becomes much larger. Third, hurdle-style analyses show that feedback discrepancy predicts whether a report changes, while feedback direction predicts the size and direction of movement among belief reports that change.

The predictor audit further clarifies what the belief update gate does and does not explain. Visible non-movement is not reducible to poor task performance, low AI literacy, low need for cognition, high confidence, or a simple ``no reason to update'' explanation. Instead, report movement is most consistently associated with local evidence structure: how discrepant the feedback is from the user's entering belief, whether the AI succeeds or fails, and where the trial occurs in the sequence. The participant-level never-mover comparison reinforces this point: participants whose reports never moved were not simply exposed to less discrepant feedback, nor were they clearly already calibrated. Visible non-movement should therefore be treated as part of the phenomenon that repeated-interaction studies need to explain.

\subsection{Implications for Human--AI Collaboration Research}

Our findings have direct implications for how trust and reliance calibration should be studied. Prior work often asks whether users appropriately increase or decrease trust in response to AI performance, confidence displays, explanations, or feedback~\cite{zhang2020effect,bansal2021whole,lu2021human,schemmer2022meta,ahn2024impact}. This remains the right broad question, but the present results show that it should be decomposed. A calibration failure can occur because feedback does not change the elicited belief report, because the report changes in the right direction but by too little, or because the report changes while downstream reliance does not. These are different failure modes and should not be collapsed into a single updating slope.

This matters for the design and evaluation of AI-assisted workflows. If corrective feedback repeatedly fails to move reported beliefs, then simply providing more evidence may not be enough; systems may need to test alternative feedback representations, cumulative performance summaries, or reflection prompts that make evidence more likely to produce measurable belief movement. Conversely, if reports move but reliance does not, the design problem is different: the system may need to support actionability, task-specific confidence, or clearer mapping from belief to decision. This aligns with prior work showing that explanations and confidence cues do not uniformly improve complementary human--AI performance or appropriate reliance~\cite{bansal2021whole,bucinca2021trust,schemmer2022meta,schoeffer2024explanations}, while outcome feedback can be more consequential for trust than interpretability alone~\cite{ahn2024impact}.

This pattern is consistent with several mechanisms: early feedback may anchor later judgments or create order effects in belief updating~\cite{tversky1974judgment,hogarth1992order}; repeated interaction may stabilize users' expectations about the AI~\cite{kahr2024understanding,kocielnik2019will,bansal2019beyond}; accumulated evidence may no longer cross the threshold needed to change an expressed report~\cite{manski2010rounding,kleinjans2014rounding,jakobsen2026coarse}; and repeated elicitation may increase satisficing or response fatigue~\cite{krosnick1991response,galesic2009effects}.

\subsection{Measurement and Model-comparison Implications}

Elicited belief reports should be treated as measurements of latent belief, not as transparent readouts. A 0--100 slider appears continuous, but participants may use it coarsely: they may round to focal values, avoid small changes, or move only when internal belief changes exceed a subjective threshold. This interpretation is consistent with latent-variable approaches to measurement~\cite{bollen1989new,borsboom2003theoretical}, work on rounded subjective probabilities~\cite{manski2010rounding,kleinjans2014rounding}, and threshold-based models for observed responses~\cite{mccullagh1980regression,agresti2013categorical}.

This measurement problem is central to interpreting repeated human--AI belief data. If many reports are flat, models that treat reported beliefs as continuously moving latent states can mischaracterize both learning and non-learning. Conversely, models that classify flat reports as pure inertia can overstate non-responsiveness. The safer interpretation is that observed non-movement identifies a report-level phenomenon: the elicited report did not change, while the underlying belief may have remained fixed or changed without crossing the reporting threshold.

The trajectory-level model comparison should be read in the same spirit. Model labels are useful descriptive motifs, not definitive participant types. A \texttt{no\_update} winner indicates that the observed report path is best summarized by non-movement under the candidate set; it does not prove that the participant's latent belief never changed. Similarly, asymmetric, threshold, sticky, or coarse-reporting labels suggest plausible patterns in reported trajectories, not uniquely identified cognitive mechanisms. This caution is consistent with model-recovery and falsification arguments in computational cognitive modeling~\cite{palminteri2017importance,wilson2019ten}. The strongest conclusion is therefore not that one model family definitively explains each participant, but that a single continuous Bayesian-updating account is too narrow once non-movement, asymmetry, thresholding, and coarse reporting are admitted as alternatives.

Future work should model the reporting process explicitly. A natural next step is a hierarchical hurdle model with three layers: an update gate governing whether evidence changes latent or expressed belief, a conditional update law governing movement direction and magnitude, and an observation model mapping latent belief into rounded, interval-censored, or thresholded reports. Such a model would directly address the two empirical signatures documented here: a large point mass at zero and substantial evidence of coarse or thresholded reporting.

\subsection{Limitations and Future Work}

It is worth noting that several limitations follow from the reanalysis design. First, the dataset was not originally designed to identify latent belief dynamics separately from reporting behavior. The analyses show that reports often do not move, but they cannot determine whether flat reports reflect true latent non-updating, latent movement hidden by coarse reporting, satisficing, or insufficient incentive to report small changes.

Second, the within-task panels are short. Ten trials per task are enough to reveal the mass of non-movement and motivate the belief update gate decomposition, but not enough to assign fine-grained cognitive mechanisms with high certainty. Longer repeated-interaction studies, stronger manipulations of feedback discrepancy, and richer process traces would make it easier to distinguish genuine inertia from thresholded expression.

Third, the analyses are descriptive rather than causal. The predictor audit shows that local evidence structure is associated with report movement, while many participant-level covariates do not explain inertial classification. These associations do not prove that changing feedback discrepancy, task order, or confidence would causally open the belief update gate. Future experiments should manipulate feedback salience, cumulative performance summaries, reporting incentives, and elicitation formats to test which interventions actually change belief movement.

Finally, the relationship between belief movement and reliance behavior deserves more direct modeling. The present paper includes descriptive delegation checks, but it does not jointly estimate belief updating, report measurement, and delegation choice. A stronger future model would do so directly, allowing researchers to ask when non-moving reports predict persistent reliance, when conditional belief movement translates into behavioral change, and when reliance changes without explicit belief movement.

%% file: sections/06_conclusion.tex
\section{Conclusions}

Repeated human--AI interaction is often explored through the lens of the extent to which people update their beliefs relative to a Bayesian benchmark. This reanalysis shows that this question is incomplete. In a balanced panel of 240 participants and 7{,}200 trials, reported beliefs frequently do not visibly move; when movement is separated from non-movement, the apparent conservatism of the pooled updating slope is substantially reduced. We introduce the \textit{`belief update gate'} framing to capture this distinction: calibration analyses should first ask whether feedback produces measurable movement in reported beliefs, and only then ask how strongly those reports move conditional on movement.

This does not contradict the original pooled belief-updating result; it decomposes it. Flat reports may reflect genuine cognitive inertia, coarse slider use, rounded probabilities, thresholds for expression, or other measurement features. Treating elicited beliefs as a single continuous updating process can therefore obscure the difference between visible non-movement, conditional learning, and coarsened reporting. Repeated AI workflows need measurement and design strategies that diagnose the belief update gate, not only average calibration.

%% file: sections/07_appendix.tex
\appendix

\section{Summary of Key Results}

Table \ref{tab:key-results} presents an overview of the main results of our work in this paper.

\begin{table}[htbp]
\caption{Summary of key results supporting the update-gate account.}
\label{tab:key-results}
\scalebox{.9}{
\begin{tabularx}{\linewidth}{
  >{\raggedright\arraybackslash}p{0.29\linewidth}
  >{\raggedright\arraybackslash}p{0.26\linewidth}
  >{\raggedright\arraybackslash}X
}
\toprule
\textbf{Diagnostic} & \textbf{Estimate} & \textbf{Interpretation} \\
\midrule
Exact zero movement 
& 67.3\% of 7{,}200 trials 
& Most trial-level belief reports do not visibly move. \\

Near-zero movement 
& 76.4\% of trial-level changes are \(<5\) percentage points 
& The distribution is concentrated at zero and small changes. \\

Slope decomposition 
& 0.494 overall; 0.949 on moved rows 
& Pooled conservatism partly reflects mixing non-movement with active movement. \\

Update gate 
& Absolute feedback gap: OR \(=3.79\), \(p=5.88\times10^{-24}\)
& Larger discrepancy between AI feedback and lagged belief predicts higher odds that the report moves. \\

Conditional movement 
& \(\hat{\beta}=.192\), \(p=2.80\times10^{-87}\) 
& Once reports move, signed feedback gap predicts direction and magnitude. \\

Trajectory comparison 
& \texttt{no\_update}: 35.5\%; \texttt{good\_bad\_news}: 26.1\%
& \texttt{no\_update} is the modal trajectory motif; asymmetry is the strongest active-updating competitor. \\

Predictor audit 
& Confidence, accuracy, and pre-survey traits not significant predictors
& \texttt{no\_update} classification is not strongly associated with the measured user traits. \\
\bottomrule
\end{tabularx}}
\end{table}

\section{Model Fitting and Selection}
\label{app:model-fit}
Each candidate model is fit separately to each participant-task trajectory. For grid-based models, we search the prespecified parameter grid and choose the parameter setting that minimizes the squared error between predicted and observed post-feedback belief reports. For conditional update-rule checks, the relevant linear slopes are estimated by closed-form least squares, with a grid over changepoint locations where needed. Predictions are clipped to the valid probability range.


For each participant-task and model, we record sum of squared error, root mean squared error, AIC, AICc, and BIC. The main results emphasize BIC because it is conservative in short panels. Because BIC penalizes model complexity, the parameter-free \texttt{no\_update} model may be favored when fit differences are small in these 10-trial trajectories. We therefore interpret BIC winner counts alongside AICc, model-recovery, absolute-fit, and source-panel sensitivity analyses rather than as definitive model assignments (Appendices~\ref{app:source-panel-robustness} and~\ref{app:absolute-fit}).


\section{Original Task Interface}
\label{app:interface}
\begin{figure*}[!ht]
\centering
\begin{minipage}{0.32\textwidth}
    \centering
    \includegraphics[width=\linewidth]{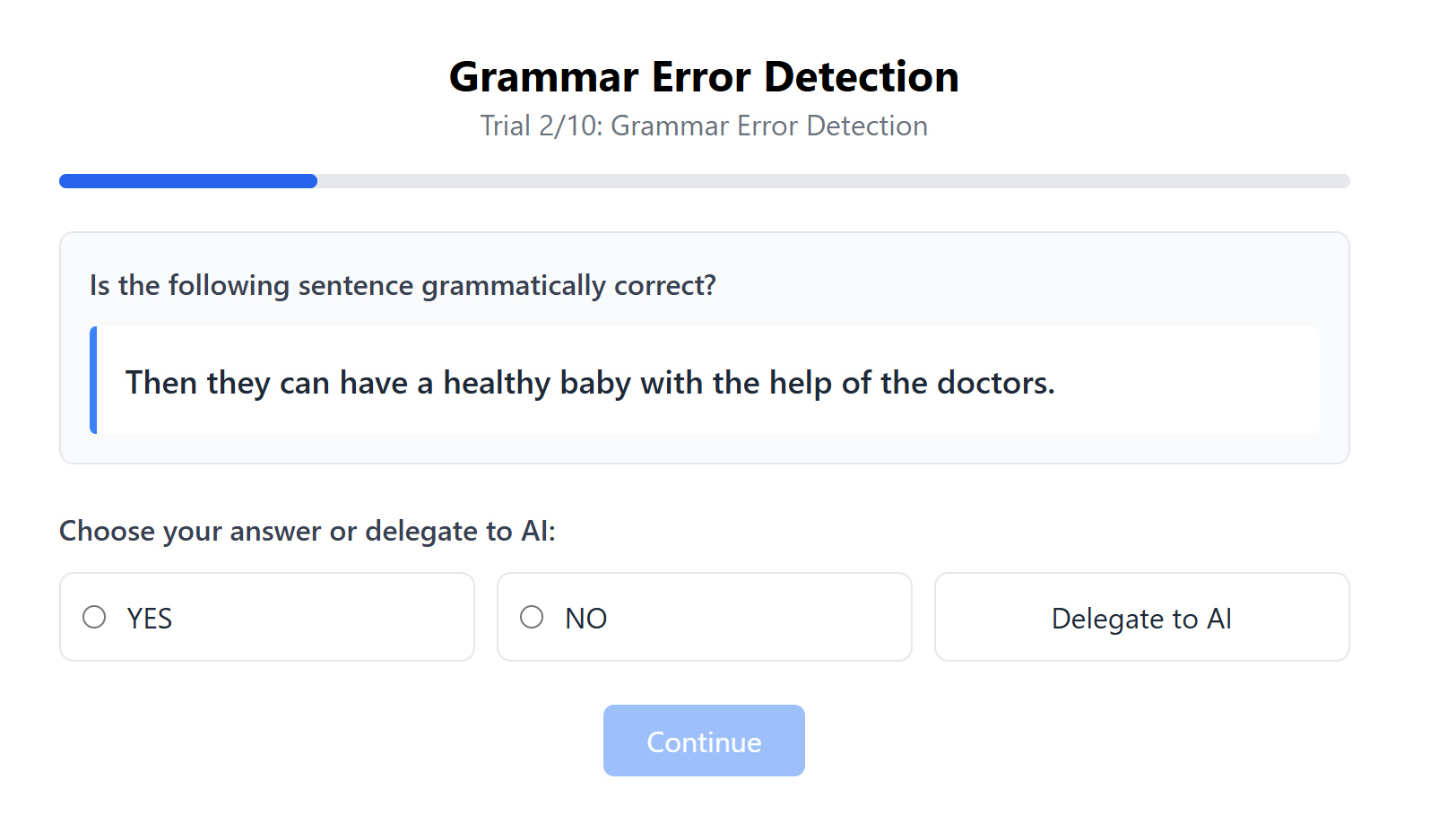}
    \caption*{(a) Grammar Error Detection}
\end{minipage}\hfill
\begin{minipage}{0.32\textwidth}
    \centering
    \includegraphics[width=\linewidth]{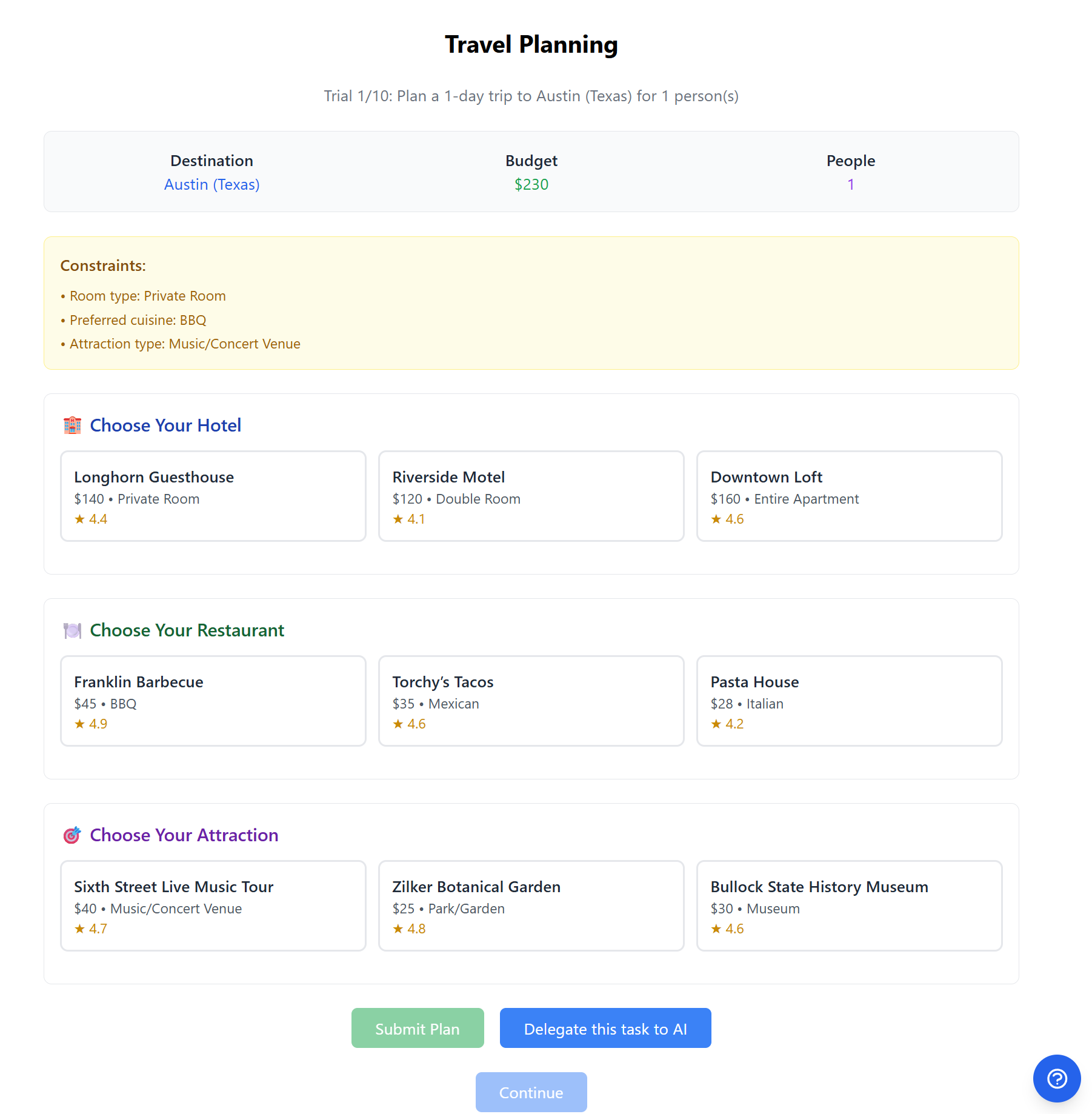}
    \caption*{(b) Travel Planning}
\end{minipage}\hfill
\begin{minipage}{0.32\textwidth}
    \centering
    \includegraphics[width=\linewidth]{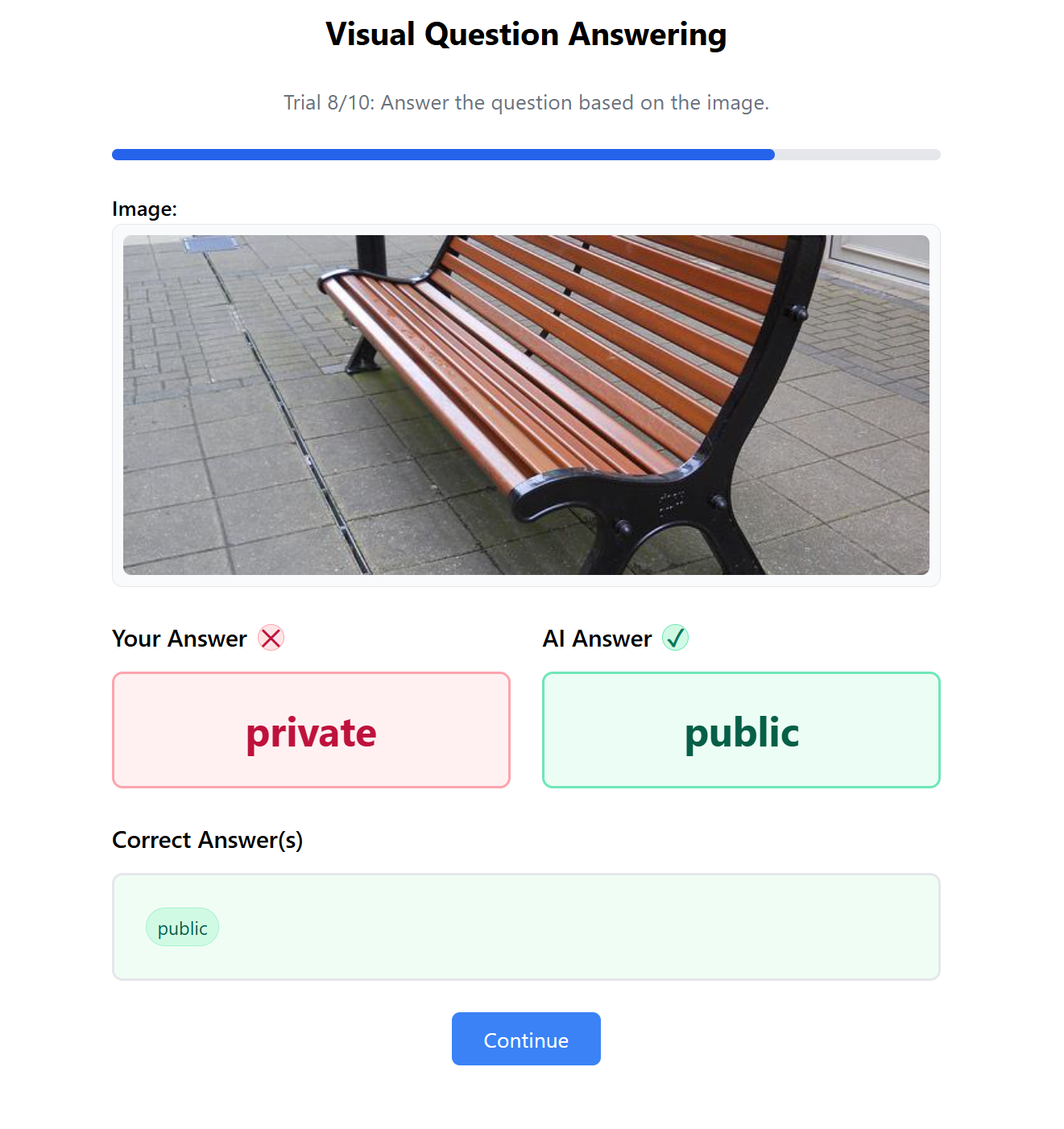}
    \caption*{(c) Visual Question Answering}
\end{minipage}
\caption{Screenshots of the three experimental tasks from the original paper \cite{biswas2026belief}. Participants completed (a) Grammar error detection, (b) travel planning, and (c) visual question answering tasks in a randomized order.}
\label{fig:task_screens}
\end{figure*}

The task interfaces that participants used for the three tasks in the original work by \citet{biswas2026belief} are presented in Figure~\ref{fig:task_screens}.

\section{Model Definitions and Robustness}
\label{app:model-robustness}

\subsection{Candidate Model Equations}
\label{app:model-equations}

This appendix records the implemented model slate. Let \(b_0\) be the
participant-task prior, \(n_0\) the inferred prior strength, \(f_t\in\{0,1\}\)
the AI correctness outcome, and \(S_t=\sum_{s=1}^{t} f_s\).

\paragraph{Standard Bayes.}
\[
  \hat b_t=\frac{b_0 n_0 + S_t}{n_0+t}.
\]

\paragraph{Divisible weighted Bayes.}
\[
  \hat b_t=\frac{b_0 n_0 + \lambda S_t}{n_0+\lambda t}.
\]
Here \(\lambda=1\) recovers the standard Bayesian benchmark, \(\lambda<1\)
indicates global evidence underweighting, and \(\lambda=0\) collapses to no
evidence accumulation.

\paragraph{Discounted Bayes.}
The discounted models replace cumulative successes and trials by exponentially
discounted quantities:
\[
  S_t^\gamma=\gamma S_{t-1}^\gamma + f_t, \qquad
  T_t^\gamma=\gamma T_{t-1}^\gamma + 1,
\]
\[
  \hat b_t=\frac{b_0 n_0+\lambda S_t^\gamma}
                 {n_0+\lambda T_t^\gamma}.
\]
The \texttt{discounted\_bayes\_gamma} model fixes \(\lambda=1\), while
\texttt{discounted\_weighted\_bayes} estimates both \(\lambda\) and \(\gamma\).

\paragraph{Coarse weighted Bayes.}
\[
  \hat b_t = Q_w\left(
  \frac{b_0 n_0+\lambda S_t}{n_0+\lambda t}
  \right),
\]
where \(Q_w(x)=\mathrm{round}(x/w)w\) clips the rounded report to \([0,1]\).

\paragraph{Sticky weighted Bayes.}
Let \(p_t^\lambda\) be the weighted Bayesian latent posterior. Reports evolve as
\[
  \hat b_t = s\hat b_{t-1} + (1-s)p_t^\lambda,
\]
where \(s\in[0,1]\) is report stickiness.

\paragraph{Threshold Bayes.}
Let \(p_t^B\) be the standard Bayesian posterior. The report updates only when
the Bayesian-implied move is large enough:
\[
  \hat b_t =
  \begin{cases}
  p_t^B, & |p_t^B-\hat b_{t-1}|\ge \tau,\\
  \hat b_{t-1}, & |p_t^B-\hat b_{t-1}|<\tau.
  \end{cases}
\]

\paragraph{Prediction-error learning.}
\[
  \hat b_t = \hat b_{t-1} + \alpha(f_t-\hat b_{t-1}).
\]

\paragraph{Good/bad-news learning.}
\[
  \hat b_t = \hat b_{t-1} + \alpha_t(f_t-\hat b_{t-1}),
  \qquad
  \alpha_t =
  \begin{cases}
    \alpha_{\mathrm{good}}, & f_t-\hat b_{t-1}\ge 0,\\
    \alpha_{\mathrm{bad}}, & f_t-\hat b_{t-1}<0.
  \end{cases}
\]

\paragraph{Confirmatory misperception.}
This model lets belief-incongruent evidence be softened before it enters the
Bayesian-style accumulator. Let \(\tilde f_t\) denote the perceived feedback
signal:
\[
  \tilde f_t =
  \begin{cases}
    q, & \hat b_{t-1}\ge .5 \ \mathrm{and}\ f_t=0,\\
    1-q, & \hat b_{t-1}< .5 \ \mathrm{and}\ f_t=1,\\
    f_t, & \mathrm{otherwise}.
  \end{cases}
\]
The posterior accumulator then updates as
\[
  \alpha_t=\alpha_{t-1}+\lambda \tilde f_t,\qquad
  \beta_t=\beta_{t-1}+\lambda(1-\tilde f_t),\qquad
  \hat b_t=\frac{\alpha_t}{\alpha_t+\beta_t}.
\]
Here \(q\in[0,.5]\) controls how strongly incongruent evidence is pulled toward
neutrality, and \(\lambda\) controls evidence weight.

\paragraph{Anchoring to prior.}
Let \(p_t^B\) be the standard Bayesian posterior. Anchoring scales the movement
away from the task-opening belief:
\[
  \hat b_t = b_0 + a(p_t^B-b_0).
\]
Values \(a<1\) indicate conservative movement toward the Bayesian posterior,
while \(a>1\) permits amplified movement.

\paragraph{Sublinear sample-size distortion.}
This model replaces the objective number of observed trials with an effective
sample size \(t^\gamma\):
\[
  \hat b_t =
  \frac{b_0 n_0 + t^\gamma \bar f_t}{n_0+t^\gamma},
  \qquad
  \bar f_t=\frac{1}{t}\sum_{s=1}^{t} f_s.
\]
When \(\gamma=1\), the model recovers the standard Bayesian benchmark; values
\(\gamma<1\) make evidence accumulate sublinearly.

\paragraph{Linear models.}
The linear models predict \(\Delta b_t\) from transformations of the normative
Bayesian step. \texttt{partial\_step} uses a single slope on the normative step;
\texttt{t1\_spike} estimates a one-time trial-1 shift; \texttt{asymmetric}
uses separate slopes for positive and negative normative steps; \texttt{recency}
uses the gap between lagged running AI accuracy and current belief; and
\texttt{changepoint} estimates separate early and late slopes around a fitted
cutpoint \(\tau\in\{1,\ldots,9\}\).

\subsection{Implemented Model Slate}
\label{app:model-slate}

\begin{table*}[ht]
\centering
\caption{Full model slate used in the all-model extension run.}
\label{tab:model-slate}
\begin{tabular}{lll}
\toprule
Model & Family & Parameters \\
\midrule
\texttt{no\_update} & inertia & none \\
\texttt{standard\_bayes} & Bayesian benchmark & none; \(\lambda=1\) fixed \\
\texttt{divisible\_weighted\_bayes} & weighted Bayes & \(\lambda\) \\
\texttt{discounted\_bayes\_gamma} & recency Bayes & \(\gamma\) \\
\texttt{discounted\_weighted\_bayes} & recency weighted Bayes & \(\lambda,\gamma\) \\
\texttt{coarse\_weighted\_bayes} & coarse reports & \(\lambda,w\) \\
\texttt{sticky\_weighted\_bayes} & response expression & \(\lambda,s\) \\
\texttt{threshold\_bayes} & inattention/threshold & \(\tau\) \\
\texttt{rescorla\_wagner} & prediction error & \(\alpha\) \\
\texttt{good\_bad\_news} & asymmetric learning & \(\alpha_{\mathrm{good}},\alpha_{\mathrm{bad}}\) \\
\texttt{confirmatory\_misperception} & confirmation & \(\lambda,q\) \\
\texttt{anchoring\_to\_prior} & anchoring & adjustment \\
\texttt{sublinear\_sample\_size} & sample-size distortion & \(\gamma\) \\
\texttt{partial\_step} & linear normative step & intercept, slope \\
\texttt{t1\_spike} & first impression & intercept, trial-1 shift \\
\texttt{asymmetric} & signed normative step & intercept, positive slope, negative slope \\
\texttt{recency} & running accuracy heuristic & intercept, slope \\
\texttt{changepoint} & phase change & intercept, early slope, late slope, \(\tau\) \\
\bottomrule
\end{tabular}
\end{table*}

\subsection{All-Candidate Winner Counts}
\label{app:all-candidate-winners}

Table~\ref{tab:all-candidate-winners} reports the all-candidate BIC winners,
including conditional update-rule diagnostics. These models are useful for
describing how observed reports move conditional on the current report, but they
are not treated as primary trajectory-generating accounts in the main text.

\begin{table}[ht]
\centering
\caption{Canonical BIC winner counts in the all-candidate model comparison.}
\label{tab:all-candidate-winners}
\scalebox{0.7}{
\begin{tabular}{lrr}
\toprule
Model & Wins & Share \\
\midrule
\texttt{no\_update} & 223 & 34.0\% \\
\texttt{changepoint} & 101 & 15.4\% \\
\texttt{good\_bad\_news} & 88 & 13.4\% \\
\texttt{t1\_spike} & 31 & 4.7\% \\
\texttt{recency} & 25 & 3.8\% \\
\texttt{asymmetric} & 21 & 3.2\% \\
\texttt{partial\_step} & 19 & 2.9\% \\
\texttt{rescorla\_wagner} & 18 & 2.7\% \\
\texttt{discounted\_weighted\_bayes} & 17 & 2.6\% \\
\texttt{sticky\_weighted\_bayes} & 17 & 2.6\% \\
\texttt{threshold\_bayes} & 17 & 2.6\% \\
\texttt{coarse\_weighted\_bayes} & 17 & 2.6\% \\
\texttt{confirmatory\_misperception} & 14 & 2.1\% \\
\texttt{anchoring\_to\_prior} & 14 & 2.1\% \\
\texttt{discounted\_bayes\_gamma} & 13 & 2.0\% \\
\texttt{divisible\_weighted\_bayes} & 10 & 1.5\% \\
\texttt{sublinear\_sample\_size} & 6 & 0.9\% \\
\texttt{standard\_bayes} & 5 & 0.8\% \\
\bottomrule
\end{tabular}}
\end{table}

\subsection{Primary-Generative Winner Counts}
\label{app:primary-generative-winners}
\begin{table}[t]
\caption{Canonical BIC winner counts among primary trajectory-generating models.}
\label{tab:winners}
\begin{tabular}{lrr}
\toprule
Model & Wins & Share \\
\midrule
\texttt{no\_update} & 233 & 35.5\% \\
\texttt{good\_bad\_news} & 171 & 26.1\% \\
\texttt{threshold\_bayes} & 33 & 5.0\% \\
\texttt{confirmatory\_misperception} & 32 & 4.9\% \\
\texttt{coarse\_weighted\_bayes} & 30 & 4.6\% \\
\texttt{anchoring\_to\_prior} & 29 & 4.4\% \\
\texttt{rescorla\_wagner} & 28 & 4.3\% \\
\texttt{sticky\_weighted\_bayes} & 25 & 3.8\% \\
\texttt{discounted\_weighted\_bayes} & 24 & 3.7\% \\
\texttt{divisible\_weighted\_bayes} & 18 & 2.7\% \\
\texttt{discounted\_bayes\_gamma} & 15 & 2.3\% \\
\texttt{sublinear\_sample\_size} & 12 & 1.8\% \\
\texttt{standard\_bayes} & 6 & 0.9\% \\
\bottomrule
\end{tabular}
\end{table}

\subsection{Source-Panel Robustness}
\label{app:source-panel-robustness}

Table~\ref{tab:source-panel-bic} reports the three most common all-candidate
BIC winners across source-panel assumptions. Every row uses the same 18-model
candidate slate; only the rule used to construct the source panel changes. The
exact number of classifiable participant-task trajectories varies because
\(n_0\) identification depends on the counterfactual-prior quality rule. The
\texttt{hybrid\_S10} row therefore reproduces the canonical all-candidate counts
in Table~\ref{tab:all-candidate-winners}.

\begin{table}[ht]
\centering
\caption{All-candidate BIC winner counts under alternative source-panel assumptions.}
\label{tab:source-panel-bic}
\scalebox{.72}{
\begin{tabular}{lrrrr}
\toprule
Panel & Trajectories & \texttt{no\_update} & \texttt{changepoint} & \texttt{good\_bad\_news} \\
\midrule
\texttt{strict} & 343 & 91 (26.5\%) & 58 (16.9\%) & 48 (14.0\%) \\
\texttt{lenient} & 567 & 175 (30.9\%) & 85 (15.0\%) & 80 (14.1\%) \\
\texttt{hybrid\_S5} & 656 & 223 (34.0\%) & 97 (14.8\%) & 87 (13.3\%) \\
\texttt{hybrid\_S10} (Canonical) & 656 & 223 (34.0\%) & 101 (15.4\%) & 88 (13.4\%) \\
\texttt{hybrid\_S20} & 656 & 223 (34.0\%) & 96 (14.6\%) & 94 (14.3\%) \\
\bottomrule
\end{tabular}}
\end{table}

\subsection{AICc Active-Updater Winners}
\label{app:active-aicc}

Among canonical \texttt{hybrid\_S10} trajectories with at least two nonzero
updates, the leading AICc winners were:
\[
\begin{array}{lr}
\texttt{good\_bad\_news}: & 86/380=22.6\%,\\
\texttt{partial\_step}: & 39/380=10.3\%,\\
\texttt{changepoint}: & 30/380=7.9\%,\\
\texttt{rescorla\_wagner}: & 28/380=7.4\%,\\
\texttt{recency}: & 24/380=6.3\%,\\
\texttt{no\_update}: & 10/380=2.6\%.
\end{array}
\]




\subsection{Update-Gate Threshold Sensitivity}
\label{app:update-gate-thresholds}
\begin{figure}[t]
    \centering
    \includegraphics[width=\linewidth]{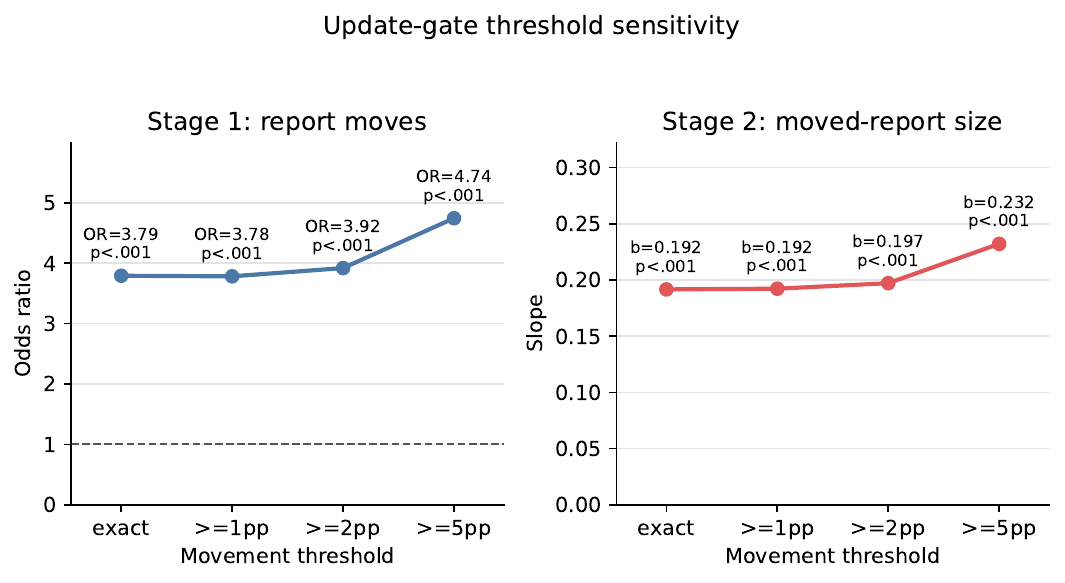}
    \caption{Hurdle-style update-gate effects across movement thresholds. The first-stage model shows that larger absolute feedback gaps increase the odds that a report visibly moves; the second-stage model shows that signed feedback gaps predict the size and direction of moved reports. All annotated effects are significant at \(p<.001\), with exact \(p\)-values reported in the text.}
    \Description{Two-panel line chart. The left panel shows first-stage odds ratios for report movement across exact, one percentage point, two percentage point, and five percentage point movement thresholds, with all labels showing p less than .001. The right panel shows second-stage signed-gap slopes across the same thresholds, again with all labels showing p less than .001.}
    \label{fig:update-gate-threshold-effects}
\end{figure}

Table~\ref{tab:update-gate-thresholds} repeats the two-stage diagnostic across
movement thresholds. The first-stage model predicts whether the report moves by
at least the threshold; the second-stage model restricts to moved rows and
predicts signed belief movement. Both stages include lagged belief, lagged
self-confidence, task domain, trial ordinal, and updater-family controls, with
participant-clustered standard errors.

\begin{table*}[ht]
\centering
\caption{Update-gate threshold sensitivity.}
\label{tab:update-gate-thresholds}
\scalebox{0.86}{
\begin{tabular}{lrrrrrrrr}
\toprule
Threshold & Moved share & Gate coef. & Gate $z$ & Gate $p$ & Cond. coef. & Cond. $t$ & Cond. $p$ & Cond. rows \\
\midrule
Exact & .327 & 1.332 & 10.09 & \(5.88{\times}10^{-24}\) & .192 & 19.80 & \(2.80{\times}10^{-87}\) & 2{,}352 \\
\(\ge\)1pp & .326 & 1.330 & 10.09 & \(5.99{\times}10^{-24}\) & .192 & 19.83 & \(1.51{\times}10^{-87}\) & 2{,}346 \\
\(\ge\)2pp & .315 & 1.366 & 10.61 & \(2.69{\times}10^{-26}\) & .197 & 20.41 & \(1.26{\times}10^{-92}\) & 2{,}265 \\
\(\ge\)5pp & .236 & 1.557 & 11.18 & \(5.30{\times}10^{-29}\) & .232 & 22.51 & \(2.99{\times}10^{-112}\) & 1{,}700 \\
\bottomrule
\end{tabular}}
\end{table*}

\subsection{Absolute-Fit Diagnostics}
\label{app:absolute-fit}

The absolute-fit summary compared fitted trajectories to observed trajectories
on several descriptive signatures in the classifiable model-comparison panel.
Three examples are particularly diagnostic:
\begin{itemize}
  \item The observed zero-update rate was 0.667. \texttt{t1\_spike} predicted
  0.647, \texttt{coarse\_weighted\_bayes} predicted 0.672, \texttt{no\_update}
  predicted 1.000, and \texttt{standard\_bayes} predicted 0.049.
  \item The observed mean final shift was \(-0.049\). \texttt{changepoint}
  predicted \(-0.049\), \texttt{asymmetric} predicted \(-0.047\),
  \texttt{partial\_step} predicted \(-0.044\), and \texttt{good\_bad\_news}
  predicted \(-0.045\).
  \item The observed mean absolute update was 0.040. \texttt{t1\_spike}
  predicted 0.038, \texttt{changepoint} predicted 0.043, and \texttt{recency}
  predicted 0.040.
\end{itemize}

\subsection{Predictor Audit for \texttt{no\_update} Classification and Visible Movement}
\label{app:inertia-predictor-audit}

Table~\ref{tab:inertia-predictor-audit} reports selected terms from the
auxiliary predictor audit. The task-level outcome is whether \texttt{no\_update}
is the BIC-winning model for a participant-task trajectory. The trial-level
outcome is whether the reported belief visibly moves on a trial.

\begin{table*}[ht]
\centering
\caption{Selected predictor-audit results for \texttt{no\_update} classification and visible belief movement.}
\label{tab:inertia-predictor-audit}
\scalebox{0.84}{
\begin{tabular}{lllrrr}
\toprule
Level & Model & Predictor & Odds ratio & Test statistic & \(p\) \\
\midrule
Task & Full pre-survey & Opening belief & 1.15 & \(z=1.25\) & .211 \\
Task & Full pre-survey & Mean confidence & 1.13 & \(z=1.05\) & .293 \\
Task & Full pre-survey & Final user accuracy & 1.00 & \(z=-0.00\) & .998 \\
Task & Full pre-survey & First AI feedback & 1.13 & \(z=1.25\) & .210 \\
Task & Full pre-survey & AI literacy & 1.10 & \(z=.75\) & .455 \\
Task & Full pre-survey & Need for cognition & .87 & \(z=-1.22\) & .223 \\
Task & Full pre-survey & Trust in automation propensity & 1.05 & \(z=.41\) & .680 \\
Task & Full pre-survey & Task domain & -- & \(\chi^2=0.42\) & .812 \\
Task & Full pre-survey & Task position & -- & \(\chi^2=7.05\) & .029 \\
\midrule
Trial & Full pre-survey & Absolute feedback gap & 1.32 & \(z=7.05\) & \(1.81{\times}10^{-12}\) \\
Trial & Full pre-survey & Lagged belief & .72 & \(z=-4.49\) & \(7.29{\times}10^{-6}\) \\
Trial & Full pre-survey & Lagged confidence & .97 & \(z=-.44\) & .658 \\
Trial & Full pre-survey & User was correct & 1.28 & \(z=6.18\) & \(6.26{\times}10^{-10}\) \\
Trial & Feedback-valence check & AI feedback correct & .88 & \(z=-3.36\) & .00077 \\
Trial & Full pre-survey & Task domain & -- & \(\chi^2=28.39\) & \(6.85{\times}10^{-7}\) \\
Trial & Full pre-survey & Trial ordinal & -- & \(\chi^2=38.65\) & \(1.33{\times}10^{-5}\) \\
\bottomrule
\end{tabular}}
\end{table*}

\subsection{Cross-Task Consistency of \texttt{no\_update} Classification}
\label{app:inertia-consistency}

This diagnostic asks whether participant-task blocks classified as
\texttt{no\_update} are stable participant tendencies or task-specific episodes.
Table~\ref{tab:inertia-consistency} summarizes the participant-level distribution
and the current-task \texttt{no\_update} classification rate as a function of how
many of the participant's other two task blocks were also classified as
\texttt{no\_update}.

\begin{table*}[ht]
\centering
\caption{Cross-task consistency of \texttt{no\_update} trajectory classification.}
\label{tab:inertia-consistency}
\scalebox{0.78}{
\begin{tabular}{lrr}
\toprule
Diagnostic & Count/share & Interpretation \\
\midrule
0 \texttt{no\_update} tasks & 125 participants; 52.1\% & No task block classified as \texttt{no\_update} \\
1 \texttt{no\_update} task & 45 participants; 18.8\% & Classification occurs in one task only \\
2 \texttt{no\_update} tasks & 28 participants; 11.7\% & Classification occurs in two tasks \\
3 \texttt{no\_update} tasks & 42 participants; 17.5\% & Classification occurs across all three tasks \\
\midrule
0 other \texttt{no\_update} tasks & 10.7\% current classification & Low current-block classification rate \\
1 other \texttt{no\_update} task & 38.4\% current classification & Moderate current-block classification rate \\
2 other \texttt{no\_update} tasks & 81.8\% current classification & High current-block classification rate \\
\bottomrule
\end{tabular}}
\end{table*}

In a participant-task logistic model controlling for task domain, task position,
opening belief, mean confidence, and final user accuracy, the standardized count
of \texttt{no\_update} classifications in the participant's other two task blocks
strongly predicts current-task \texttt{no\_update} classification (odds ratio
\(=4.46\), \(z=12.00\), \(p=3.52\times10^{-33}\), \(n=720\)). By contrast, a
sequential check asking whether the immediately previous task was classified as
\texttt{no\_update} does not show a positive carryover effect (odds ratio \(=.74\),
\(p=.235\), \(n=320\)). Thus \texttt{no\_update} classification shows some
cross-task consistency, but it is neither universal nor a simple one-step carryover
pattern.

Table~\ref{tab:slope-bootstrap} reports participant-bootstrap confidence
intervals for the slope decomposition. The intervals resample participants,
then recompute the within-participant-task slope in each subset.

\begin{table*}[ht]
\centering
\caption{Participant-bootstrap confidence intervals for slope decomposition.}
\label{tab:slope-bootstrap}
\scalebox{0.78}{
\begin{tabular}{lrr}
\toprule
Subset & Slope & 95\% CI \\
\midrule
All rows & .494 & [.405,.586] \\
Exclude \texttt{no\_update} winners & .682 & [.590,.792] \\
\(\ge\)1 nonzero-update trajectory & .669 & [.577,.780] \\
\(\ge\)2 nonzero-update trajectory & .719 & [.617,.835] \\
Nonzero rows only & .949 & [.831,1.081] \\
Nonzero minus all & .455 & [.370,.543] \\
\bottomrule
\end{tabular}}
\end{table*}